# Searching and Establishment of S-P-O Relationships for Linked RDF Graphs : An Adaptive Approach

Ayan Chakraborty
Dept. of Computer Science Engineering
Techno India College of Technology
Kolkata- 700156, India
Web Intelligence &
Distributed Computing Research Lab
Golf Green, Kolkata: 700095, India
Email: achakraborty.tict@gmail.com

Shiladitya Munshi
Dept. of Information Technology
Meghnad Saha Institute of Technology
Kolkata- 700150, India
Web Intelligence &
Distributed Computing Research Lab
Golf Green, Kolkata: 700095, India
Email: shiladitya.munshi@yahoo.com

Debajyoti Mukhopadhyay
Department of Information Technology
Maharastra Institute of Technology
Pune: 411038, India
Web Intelligence &
Distributed Computing Research Lab
Golf Green, Kolkata: 700095, India
Email: debajyoti.mukhopadhyay@gmail.com

*Abstract*—In the coming era of semantic web linked data analysis is a very burning issue for efficient searching and retrieval of information. One way of establishing this link is to implement subject-predicate-object relationship through Set Theory approach which is already done in our previous work. For analyzing inter-relationship between two RDF Graphs, RDF-Schema (RDFS) should also be taken care of. In the present paper, an adaptive combination rule based framework has been proposed for establishment of S-P-O (Subject-Predicate-Object) relationship and RDF Graph searching is reported. Hence the identification of criteria for inter-relationship of RDF Graphs opens up new road in semantic search.

## I. INTRODUCTION

The Resource Description Framework (RDF) is a family of World Wide Web Consortium (W3C) specifications originally designed as a metadata model. As per W3C study, it supports a generalized method for conceptual description or modeling of information which uses various types of syntactical formats to be implemented in web resources.In semantic environment RDF represent the web data as subject-predicate-object triplet format which are basically the URIs or literals and therefore the linking structure of the Web is redefined. This triple form of RDF leads to the path of Graph representation which comprises of the interrelationship between more than one RDF triples using S-P-O linking. This is called as N-Triple Graph. The following N-Triples file consists of three RDF statements:

TABLE I.    N-TRIPLE REPRESENTATION OF RDF

| N-Triples | |
|---|---|
| 1 | Subject:<http://www.w3.org/2001/sw/RDFCore/ntriples/><br>Predicate:< http://purl.org/dc/elements/1.1/creator><br>Object: "Dave Beckett" |
| 2 | Subject:<http://www.w3.org/2001/sw/RDFCore/ntriples/><br>Predicate: <http://purl.org/dc/elements/1.1/creator><br>Object: "Art Barstow" |
| 3 | Subject:<http://www.w3.org/2001/sw/RDFCore/ntriples/><br>Predicate: <http://purl.org/dc/elements/1.1/publisher><br>Object: <http://www.w3.org/> |

In the following table [16] three sets of RDF Triples are represented. The relationships expressed in these triples lead to the establishment of the relation in between the resources and property values of RDF which is nothing but the Subject-Predicate-Object relations. So the representation of RDF data in triple format is known as N-Triple Graph or NTGraph. Establishment of the interrelationships in between NTGraphs and their characterization is proposed in this following framework using simple Set Theoritic approach. This will impart a great effect in the searching of RDF data in semantic environment.

When RDF Graphs are represented with set diagrams, the intersection of given two RDFSets will denote the relationship between them. The set representation of an RDF consists of three subsets: subject, predicate and object. Moreover for the categorization of these inter-relationships the impact of RDF Schema is very important.

In this paper, for the establishment and categorization of S-P-O relationships, the impact of Set theoretic approach is presented.

## II. LITERATURE REVIEW

Semantic Web is a metadata data model. In this environment data models and syntax fro the interrelations are specified by RDF[6]. This data model is consisted of URI[4] or literals. There may be some resources which are not containing any URIs or literals. They are "anonymous" resources and called as Blank Nodes[5].

Jonathan Hayes proposes a map from RDF Graphs to directed labeled graphs in his diploma thesis [1]. This proposal is thought to complement the default representation of RDF by graphs as provided by the RDF specification[2]. Even though this proposal claims to be formal and unambiguous, the fundamental limitations inherent in the modeling of RDF as directed labeled graph persist.

The graph representation of an RDF is basically the the mapping of S, P or O triplets. In this context, ontology matching [7, 8, 9,11] (ontology alignment) and instance matching are the two most-studied sub-problems of interlinking. The latter papers often refer to the process of determining whether two descriptions refer to the same real-world entity in a given domain[12,13]. Although these two problems are related, they are neither necessary nor sufficient to solve each other. Samur Araujo1, Jan Hidders1, Daniel Schwabe2, Arjen P. de Vries [3] has given solution for the instance-matching problem is composed of two phases: the selection phase and the disambiguation phase. In the selection phase, for each instance r in a dataset A, they search for instances in a target dataset B that may refer to the same entity in the real world as r, by using a literal type of matching that has a high recall but a low precision.

With the essence of those previous works, a new concept has been proposed in this paper for identification and categorization of the relationship between two RDF Graphs that represent RDF triples. This relationship is primarily driven by the subject-predicate-object relation of RDF triples and RDFS vocabularies.

### III. INTER-RELATIONSHIP BETWEEN RDF GRAPHS

In this paper RDF Graphs are represented with set diagrams where *subject*, *predicate* and *object* are three subsets. The intersection of given two RDFSets will denote the relationship between them. But existence of a blank node is critical to this point. In RDF Graph representation, a node, which is not containing any URI or literal, is called as a blank node or bnode. The resource represented by a blank node is also called an anonymous resource. By using Reification technique the blank nodes can be represented in RDF Graph model. The anonymous resource (bnode) is reified by splitting up into two triples. This concept is shown in Fig.2:

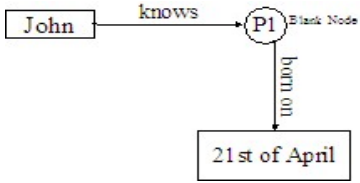

Fig. 1. Blank Node Reification

In the statement "John has a friend born on 21st of April", the expression can be represented by coupling two triples with a blank node denoting the anonymous friend of John.

John<subject> knows<predicate>   $p_1$ <object>
$p_1$<subject>   born on<predicate>   $21_{st}$ of April<object>
where $p_1$ is the bnode.

Now the blank nodes are also converted into general triple format. So, the conditions for the intersection in between these subsets are given below:

Let V be a vocabulary, T be an RDF Graph with vocab(T)⊆V and $G_{dir,label,multi}$ the set of directed, edge- and node-labeled multigraphs. We then define a map

$\delta$: RDF Graph(V)→$G_{dir,label,multi}$

as follows: $\delta(T)= (N,E,L_N,L_E)$, where
N={$n_x$:x∈Subj(T)∪Obj(T)}
$L_N(n_x)=$ x,$d_x$   if x is literal and $d_x$ is datatype identifier
x   else

E={$e_{s,p,o}$:(S,P,O)∈T}
from $(e_{s,p,o})=n_s$,   to $(e_{s,p,o})=n_o$   and $L_E(e_{s,p,o})$=P

The inter relationship between two RDF Graph is important from semantic search perspective. As mentioned in previous discussions, the semantic expressiveness of a statement can potentially be stored through RDF Graph and hence discovery of inter relationship criteria for RDF Graphs could form the basis of a formalized graph based search algorithm in the context of reservation and exploration of semantic nature of the statements or assertions.

On this background following section characterizes the criteria for establishing relationship between two RDF Graphs.

### IV. CHARACTERIZATION OF INTER-RELATIONSHIP BETWEEN TWO RDF GRAPHS

The need of characterization of RDF Graph relationship is steeply growing as more and more data are being published in Semantic Web with RDF standard. The challenge of mining data from Semantic Web mostly can be met with the proper identification of related triples *based* on semantic constructs like subject or objects etc. The network of those related RDF Graphs forms the local reference frame for the information to be searched.

On the context of present discussion following subsections characterizes RDF Graphs through a Set Theory approach [15].

#### A. Subject-Subject and Predicate-Predicate relationship

Subject-Subject and Predicate-Predicate relationship described in Fig. 2, characterizes the specific criteria of RDF Graph relationship, where two RDF Graphs T1 and $T_2$ share common subject and predicate. The significance of these criteria is that two statements are semantically equivalent from the Subject and its property perspective. The only difference exists in a point that the two statements have different values for the same properties of the same subjects. It is evident that this criteria dictates a *strong* relationship between two RDF Graphs $T_1$ and $T_2$ and between two corresponding statements as well.

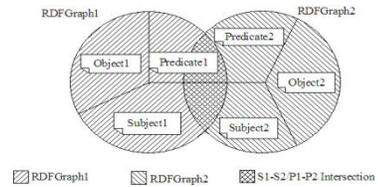

Fig. 2. Subject-Subject/Predicate-Predicate Relationship of two given RDF Graphs

Mathematically, there exists a Subject-Subject and Predicate-Predicate relationship between two RDF Graphs $T_1$ and $T_2$, if the following set theoretical expressions are all true.

$Sub(T_1) \cap Sub(T_2) = \emptyset$
$Obj(T_1) \cap Obj(T_2) = \emptyset$
$Sub(T_1) \cap Obj(T_2) = \emptyset$
$Obj(T_1) \cap Sub(T_2) = \emptyset$
$E_1 \cap E_2 = \emptyset$

Following is an example of above mentioned relationship:

T1: Subject:http://www.example.org/staffid/85740
Predicate:http://www.example.org/terms/desig
Object:http://www.example.org/dept/accountant

T2: Subject:http://www.example.org/staffid/85740
Predicate:http://www.example.org/terms/desig
Object:http://www.example.org/club/treasurer

*B. Object-Object and Predicate-Predicate relationship*

Object-Object and Predicate-Predicate relationship identifies the criteria of RDF Graph relationship, where two RDF Graphs $T_1$ and $T_2$ share common object and predicate. Two RDF Graphs related with this kind of condition, must have different *subjects* which hold same property with same values. Object-Object and Predicate-Predicate relationship has immense importance where the semantic search is based on some property and its specific values. It is evident that this criteria also dictates a *strong* relationship between two RDF Graphs $T_1$ and $T_2$ and between two corresponding statements as well.

A condition for Object-Object and Predicate-Predicate relationship is presented in Fig.3.

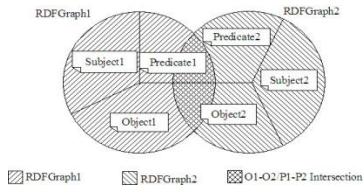

Fig. 3. Object-Object/Predicate-Predicate Relationship of two given RDF Graphs

Mathematically, there exists a Object-Object and Predicate-Predicate relationship between two RDF Graphs $T_1$ and $T_2$, if the following set theoretical expressions are all true.
$Obj(T_1) \cap Obj(T_2) = \emptyset$
$Sub(T_1) \cap Sub(T_2) = \emptyset$
$Sub(T_1) \cap Obj(T_2) = \emptyset$
$Obj(T_1) \cap Sub(T_2) = \emptyset$
$E_1 \cap E_2 = \emptyset$

Following is an example of above mentioned relationship:

T1: Subject:http://www.example.org/staffid/85740
Predicate: "published"
Object:http://www.wikipedia.com/technology/C.V.

T2: Subject:http://www.example.org/staffid/85742
Predicate: "published"
Object:http://www.wikipedia.com/technology/C.V.

*C. Subject-Predicate relationship*

Subject-Predicate relationship has a different significance and consequence than the other two types of relationships discussed above. In this case, two RDF Graphs $T_1$ and $T_2$ never share their subject, object or predicate, rather the resource described by one's subject is same as that of resource described as predicate of others. With this condition, the subject of one statement acts as a property of the other statement. The two RDF Graphs related with their Subject - Predicate relation can represent complex indirect search construct. The two statements with completely different subjects could be linked with each other through this relationship.

A condition for Subject-Predicate relationship is presented in Fig.4. Mathematically, there exists a Subject-Predicate relationship

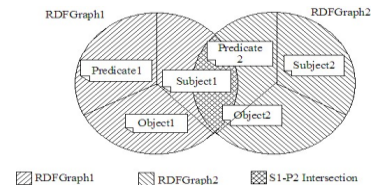

Fig. 4. Subject-Predicate Relationship of two given RDF Graphs

between two RDF Graphs $T_1$ and $T_2$, if the following set theoretical expressions are all true.
$Sub(T_1) \cap Sub(T_2) = \emptyset$
$Obj(T_1) \cap Obj(T_2) = \emptyset$
$Sub(T_1) \cap E_2 = \emptyset$
$E_1 \cap E_2 = \emptyset$

Following is an example of above mentioned relationship:

T1: Subject:http://www.example.org/staffid/85740
Predicate:http://www.example.org/terms/desig
Object:http://www.example.org/dept/accountant

T2: Subject:http://www.example.org/terms/desig
Predicate:http://www.example.org/staffid/85740
Object:http://www.example.org/club/treasurer

## V. SEARCHING RDF GRAPH SEQUENCES: INTRODUCING ADAPTIVENESS

In the above sections, how the relationships can be established and quantified among n number of RDF Graphs is proposed. These relationships can be of SS-PP, OO-PP, S-P and O-P within 1 to n-dimensional status. So the resources about any subject in WEB 3.0 will be chunked in the RDF Graphs which are inter-related using those quantified relationships as well as form a complex Graph sequences.

Now the question is that, if a query is triggered about any subject, the information about that subject is scattered in those complex RDF_Relational_Graph structures which are individually not inter-related. So the the searching will be partial. To overcome this difficulty, all the RDF_Relational_Graphs are represented as Relational_Sequences and using these sequences, *Relational_Sequences* storage has been created. From this storage using *Gen____Repetitive__Seq* and *Gen Relational_Pattern* algorithm (described in next section) a *Relational_Pattern* will be derived. As this *Relational_Pattern* is a subset of all *Relational_Sequences*, when searching information for a specific subject, using this *Relational_Pattern* all the individual SS-PP, OO-PP, S-P and O-P RDF Relational Graphs can be completely searched.

Let us consider a Set $A$ where $A = \{S^0, S^1 \cdots S^t\}$ and up to $t$ numbers of *Relational_Sequence* can be stored within Set $A$. Now, using the Adaptive Algorithm (Algorithm 1)[14] *PatternAnalyzer* will identify $k$ number of $l$ length *Relational_Pattern* out of Set$A$. The length $l$ will be tuned adaptively in between two limiting values and $k \geq 1$. These limiting or range values will clarified in the next section. Using all the identified k *number* of l *length Relational_Pattern*s, another Set $B$ will be formed.

K1 and K2 are two boundary integer values which will tune the length of the *Relational_Pattern* l. Initially the value of K1 will be assigned by 1 at t=0, otherwise for t >0 and for the next sessions the value of K1 will be equal to the value of the length of last *Relational_Pattern*. The value of integer K2 will be assigned up to the total length of the *Relational_Sequences* i.e.; n. Let us consider an integer variable x with a value within the range of K1 and K2. Identification of a *Relational_Pattern* for any specific *Relational_Sequences* storage is basically driven by the value of x. Another integer variable r has to be considered to define the number of most repetitive sequence identified for a given length x. Now in the time of computation of most repetitive sequence for a specific *Relational_Sequences* storage, many pairs of (r; x) will be generated. The ratio of each pair of (r; x) i.r.; $\lambda = r/x$ actually effects the lengthwise variation of the *Relational_Pattern* l.

### A. Algorithm: Generation of Relational Pattern for S-P-O

With the discussion of previous topics, following sections introduce the theoretical sides of an adaptive algorithm to find $k$ number of $l$ symbol length *Relational_Pattern*s.

Algorithm 1: Gen_Relational_Pattern
*Input*: (i) *Relational_Sequences* $S^t$ for $t^{th}$ session, along with
  *Relational_Sequences* storage containing $S^0, S^1 \cdots S^{t-1}$
  and $P_s^0, P_s^1 \cdots P_s^{t-1}$
  (ii) $K_1 = 1$ (if $t = 0$)  $\hat{x}^{t-1}$ (otherwise)
  (iii) $K_2 = n$ where $n$ = length of *Relational_Sequences*s

*Output*: *Relational_Pattern* $P_s^t$ for $t^{th}$ Relational_Sequences
*Step 0*: for all $x$ (where $K_1 \leq x \leq K_2$), do *Step 1* to *Step 2*
*Step 1*: call *Gen_Repetitive_Seq* subroutine (described next) to have repetitive sequence $P$ and number of repetitive sequence $r$.
*Step 2*: compute $\lambda = r/x$ and if $\lambda \leq \lambda_{min}$, $\lambda_{min} \leftarrow \lambda$
*Step 3*: designate $x$ for which $\lambda = \lambda_{min}$ as $\hat{x}^t$
*Step 4*: $P_s^t \leftarrow P$
*Step 5*: Stop.

Adaptiveness is implemented by the algorithm stated above for the identification of *Relational_Pattern* using the derived knowledge of *Relational_Sequences* storage. Although the repetitive sequence search or the most subsequent occurrence search is controlled by deterministic approach, change in the value of $K_1$ boundary in every iteration, makes the next session adaptive with the knowledge which is derived and translated via $\hat{x}^{t-1}$ to the next session for each iteration. The decreasing difference between two boundaries values of $K_1$ and $K_2$ depends on the adaptiveness of $\hat{x}^{t-1}$ and as a result the searching time of *Relational_Pattern* in between RDFs will be minimized with the progress of time.

A search for repetitive sequence within *Relational_Sequences* storage is described next. This search could be optimized (with respect to time) by employing any smarter deterministic algorithm. The time complexity of the algorithm *Gen_Repetitive_Seq* to search out a repetitive sequence of any length from the *Relational_Sequences* storage is $(r(n * p)^2)$. Hence the above algorithm requires $((K_2 - K_1)r(n * p)^2)$ time to search for desired *Relational_Pattern*, by repetitive calling of this sub-routine . So *Relational_Pattern* generation can be done in polynomial run time.

### VI. MATCHING OF RDF GRAPHS USING S-P-O RELATION

To apply Dempster Shafer Rule of Combination in two RDF Graphs, a non empty finite set $\{A\}$ has been taken, which contains the *URISequence* at RDF Graph1, i,e $S^1$ gets directly mapped with $\{A\}$. Similarly let us take another non empty finite set $\{B\}$ containing the *URISequence* at RDF Graph2, i,e $S^2$.

Next, let us consider $m$ = Mass Function of Belief or Basic Belief Assignment and hence as per Dempster Shafer Rule of Combination, the belief function of $\{A\}$ i,e $bel(A)$ is given by

$$bel(A) = \sum_{B/B \subseteq A} m(B).$$

This equation directly derives

$$m(A) = \sum_{B/B \subseteq A} (-1)^{/A-B/} bel(B).$$

Therefore $bel(B)$ can be given as

$$bel(B) = \frac{k}{\sum_{B/B \subseteq A} (-1)^{/A-B/}}$$

where $k$ is Shafer Normalization Factor and the mathematical expression is

$$k = \sum_{x_1 \cap x_2 \cdots \cap x_n = A = \varphi} (m_1(x_1)m_2(x_2) \cdots m_n(x_n))$$

Here $x_1, x_2 \cdots x_n$ are the relational scoring of $n$ number of different inter-relationships between RDF graphs within the scope of $S^1$ and $S^2$.

So $bel(b)$, defines the degree of belief of the probability of the occurrence of $URISequence$ of RDF Graph2, i.e $S^2$ in the superset $S^1$. Hence, if the value of $bel(b)$ coincides with the preset value, the S-P-O relationship will be established.

Therefore conclusively the proposed framework works within polynomial run time as both $URISequence$ generation and computation of the balue of $bel(B)$ will be done within polynomial run time.

However, the current study has identified some other types of relationships between two RDF Graphs which are not significant from the semantic search perspective. The Subject-Subject relationship suffers from the unboundedness. The absence of any common property (predicate) or its value (object) fails to limit the growth of this relationship, hence this could be considered as a *weak* relationship. Predicate-Predicate or Object-Object relationships has no intrinsic search value, because these two relationship holds an incomplete search criteria over two RDF Graphs. As a result, this is noteworthy that Subject-Subject, Object-Object and Predicate-Predicate relationship has insignificant impact over RDF Graph based semantic search.

## VII. CONCLUSION

The current study has successfully met its objective of basic characterization of inter-relationship between two RDF Graphs. Set theory expressions guided by the adaptive framework using RDFS vocabularies have been identified which could be considered as necessary and sufficient conditions for discovering relationships between a RDF Graph pair. The potential of this investigation can further be exploited with future research focusing on probabilistic quantification of these relationships and graph traversal based search algorithms within the network of such inter-related RDF Graphs.